\documentstyle[11pt,newpasp,twoside,epsf]{article}
 \def\ltsim{\lower.5ex\hbox{$\; \buildrel < \over \sim \;$}}

\def\edcomment#1{\iffalse\marginpar{\raggedright\sl#1\/}\else\relax\fi}
\reversemarginpar

\begin{document}
\title{What is the Broad Line Region?}
\author{Ari Laor}
\affil{Physics Department, Technion, Haifa 32000, Israel}

\begin{abstract}
What is the Broad Line Region (BLR) made of? What determines its location?
Why is it sometimes
missing? What controls its properties? Some recent results
and new approaches which may shed light on these issues are
briefly described.
\end{abstract}

\section{Introduction}
Imagine an Earth like civilization which developed on an isolated
planet away from any star or galaxy.
An astronomer there, faced with the spectrum of a spatially unresolved
galaxy, will deduce it is made of dense gas
``clumps'' with $n\sim 10^{17}$~cm$^{-3}$ and $T\simeq 3,000-30,000$~K.
Yet, he will have almost no idea of the true nature of galaxies.
Here, on Earth, we may be facing a similar difficulty when trying to
understand the BLR just based on its spectrum.

Photoionization calculations have reached a high
level of sophistication, providing good constraints on the
density, temperature, ionization level, chemical abundances and spatial
distribution of the gas in the BLR (e.g. Ferland, these proceedings). However,
these calculations do not
reveal the origin of the gas. Is it made of individual ``clouds"
produced by e.g. bloated stars, tidally disrupted stars,
a gravitationally unstable outer disk, or
a wind interaction with SN remnants?  Or, is it made of a
continuous flow, e.g. a wind driven off an accretion disk by
radiation pressure, magnetic pressure, or both?   What determines the
location of the BLR? Is it accretion disk physics?
Furthermore, AGN show many emission lines with
a large range in strengths and profiles, yet these emission
properties are strongly correlated, suggesting there may be just a few free
parameters which control these properties. What are these free parameters?
In this short review I will briefly describe some of the progress
made towards answering the above questions, and some potentially
useful new approaches.

\section{What is the BLR made of?}
Photoionization calculations tell us the thickness of the line
emitting layers, and their solid angle,
but they do net tell us whether the gas is composed
of many small units (e.g. bloated stellar atmospheres), a few larger
units (e.g. tidally disrupted stars, wind interaction with supernovae
remnants), or a single continuous flow (e.g. a disk outflow).

\begin{figure}
\plotfiddle{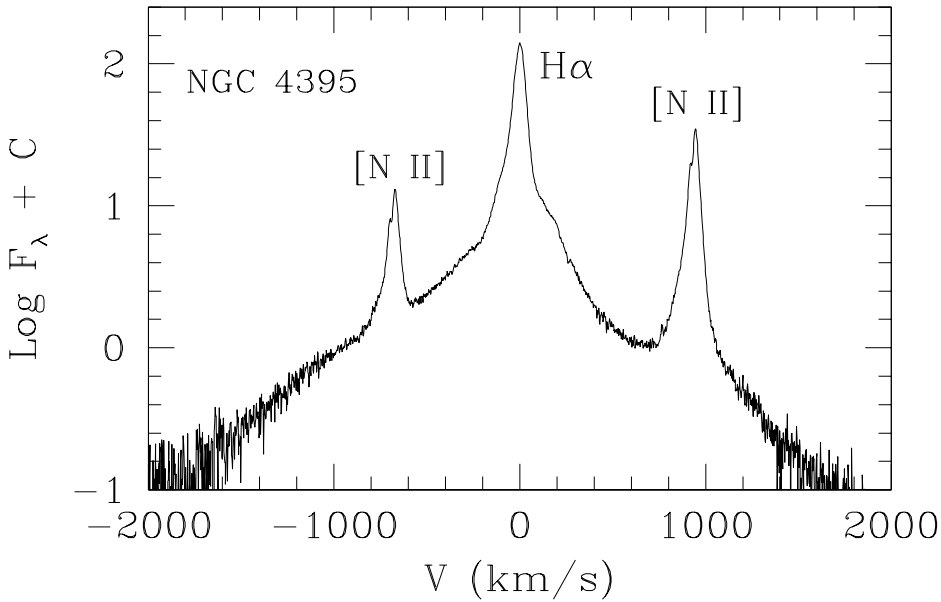}{7.5cm}{0.}{80}{80}{-240}{-320}
\caption{The Keck profile of continuum subtracted H$\alpha$ profile
in NGC~4395, the lowest luminosity Seyfert 1
galaxy. The profile remains smooth despite the small size of the
BLR (courtesy of A. Filippenko and L. Ho.)}
\end{figure}

The number of ``clouds" in the BLR,
$n_c$, and their size, $r_c$, can be constrained based on the
smoothness of the emission line profiles. The level of
fluctuations in the emission line profiles is proportional to
$1/\sqrt{n_c}$ (see Capriotti et al. 1981 for a nice simulation), assuming the
cloud velocities are randomly distributed within the line
profile. The total solid area subtended by the BLR clouds is
$\Omega_{\rm BLR}=n_c\pi r_c^2/4\pi R_{\rm BLR}^2$, where $R_{\rm BLR}$
is the BLR radius. Photoionization calculations
indicate that $\Omega_{\rm BLR}\simeq 0.1-0.2$, and thus
$r_c\simeq R_{\rm BLR}/\sqrt{n_c}$. Very high spectral resolution
and S/N observations allow a large lower limits on $n_c$, and thus a
tight upper limit on $r_c$, in particular in low luminosity
objects where $R_{\rm BLR}$ is expected to be small.

A recent study of the H$\alpha$ profile by Arav et al. (1998)
using a very high S/N Keck spectrum placed a limit of
$n_c>3\times 10^7$ in NGC~4151, where
$\nu L_{\nu}(5100$\AA$)=7\times 10^{42}$~erg~s$^{-1}$.
A similar study by Dietrich et al. (1999) in 3C~273, where
$\nu L_{\nu}(5100$\AA$)\simeq 6\times 10^{45}$~erg~s$^{-1}$,
yielded  $n_c>10^8$. In Figure~1 we show a high quality Keck spectrum
of the H$\alpha$ profile in NGC~4395, the least luminous Seyfert~1 galaxy,
where
$\nu L_{\nu}(5100$\AA$)=6\times 10^{39}$~erg~s$^{-1}$. At this luminosity
one expects $r_{\rm BLR}\simeq 10^{14}$~cm, if
$R_{\rm BLR}\propto L_{\nu}^{0.7}$ (Kaspi et al. 2000), or
$R_{\rm BLR}\simeq 6\times 10^{14}$~cm, if
$R_{\rm BLR}\propto L_{\nu}^{0.5}$. In the bloated stars scenario
(Alexander \& Netzer 1994) the size of the individual stars is
$\sim 10^{14}$~cm, thus only a handful of such stars can be accommodated
in the BLR of NGC~4395. This contrasts with the smoothness of H$\alpha$
in NGC~4395 which implies $n_c>10^6$, and thus $r_c<10^{11}-10^{12}$~cm.
The bloated stars scenario is therefore ruled out,
together with other scenarios which involve individual units
significantly larger than  $10^{12}$~cm. The smoothness of the BLR profiles
in the lowest luminosity AGN therefore suggests
that the BLR most likely originates in a single continuous flow.

\section{What determines $R_{\rm BLR}$?}

There are various models which attempt to explain the formation of the BLR based
on accretion disk models. For example, it may arise in a gravitationally
unstable outer disk,
a disk outflow resulting from a combination of disk instability and
a confining disk corona, or various other disk outflow mechanisms.
However, some of the models do not make specific predictions for $R_{\rm BLR}$,
or otherwise predict that it should be a function of the black hole mass
$M_{\rm BH}$
in addition to $L$. These predictions may be inconsistent with the
observations which indicate a rather tight relation
of the form $R_{\rm BLR}\simeq 0.01(L_{\rm bol}/10^{44})^{0.62-0.70}$~pc
(Kaspi et al. 2000; Peterson et al. 2000).

Alternatively, $R_{\rm BLR}$ may be set by dust sublimation. Dust embedded in
the ionized gas strongly suppresses line emission when the ionization parameter
$U>0.01$. The column of the photoionized
H~II layer is $\Sigma_{\rm ion}\simeq 10^{23}U$~cm$^{-2}$, and
the UV optical depth of dust is $\tau_{\rm dust}\simeq \Sigma/10^{21}$,
which implies $\tau_{\rm dust}({\rm H~II})\simeq 100U$. Thus,
when $U>0.01$ the
penetration of ionizing radiation to $\Sigma_{\rm ion}> 10^{21}$~cm$^{-2}$
is suppressed by a ``dust wall'', which converts the excess ionizing continuum to
dust IR emission. The outgoing line emission is further suppressed
due to absorption by the embedded dust.
Since all dust sublimates below $R_{\rm sub}\simeq
0.02(L_{\rm bol}/10^{44})^{0.5}$~pc, efficient line emission is
restored at $R< R_{\rm sub}$. Dust sublimation thus provides  a
{\em parameter free} mechanism
for the outer boundary of the BLR (Laor \& Draine 1993; Netzer \& Laor 1993;
Ferguson et al. 1997).

Just outside the BLR the gas cools mainly by dust
IR emission, producing the ubiquitous 3~$\mu m$ bump. Still further
out low density gas with $U<0.01$ produces the NLR emission. Further
inward from the BLR the density of $U\ltsim 1$ gas increases, which leads to
collisional destruction of the lines, and cooling mostly by atomic
continuum radiation. Thus, the dust scenario implies that reprocessing
gas is not necessarily located just at the BLR, but may be located at all radii,
and the BLR is just where the gas cools efficiently by line emission.

The dust and disk BLR scenarios imply different responses of the
BLR emission to ionizing continuum variations.
If $R_{\rm BLR}$ is set by the accretion disk structure then it
is expected to respond to variations in the disk structure,
which occur at most at the disk sound speed crossing time.
Since the sound speed in a thin accretion disk
is well below the local Keplerian speed, one expects changes
in the BLR structure to occur on timescales much longer than
the dynamical time
$t_{\rm dyn}=R_{\rm BLR}/v_{\rm BLR}$. Any other mechanism
which produces physical changes in the gas distribution
in the BLR cannot occur faster than $t_{\rm dyn}$.

In the dust bounded BLR hypothesis, reprocessing gas is present
at all radii, and changes in $R_{\rm BLR}$ occur at the dust
sublimation and reformation timescale.
The sublimation timescale just outside the BLR is
$t_{\rm subl}\sim 10^{3-5}$~s, and one thus expects that an
increase in $L$ should be accompanied by an increase in the broad
line core flux from the outer BLR
with a delay of only $\sim R_{\rm BLR}/c$. Since
$v_{\rm BLR}\simeq 0.01c$ the dust scenario allows profile
variations which are 100 times faster(!) than the fastest variability timescale
allowed by physical changes in the BLR (e.g. the disk BLR scenario).

When $L$ decreases dust can reform at smaller radii, again
maintaining the $R_{\rm BLR}\propto L^{0.5}$ relation. However,
dust condensation is a much slower processes ($t_{\rm cond}\sim 10^{\ge 7}$~s),
and since nucleation cannot start in a $\sim 10^4$~K photoionized gas,
it will take $\ge t_{\rm dyn}$ to get a fresh supply of dusty
gas into the outer BLR. Thus, if $R_{\rm BLR}$ is set by
dust sublimation then $R_{\rm BLR}(L,t)$ will show a {\em hysteresis effect},
i.e. it will follow the $R_{\rm BLR}\propto L^{0.5}$ relation when
$L$ increases, but when $L$ decreases  $R_{\rm BLR}$  will respond with a
delay of at least $t_{\rm dyn}$. In the disk scenario there should not
be any clear asymmetry in the response of  $R_{\rm BLR}$ to luminosity increases
vs. decreases. Finally, a corresponding hysteresis effect should be observed in the
response of the 3~$\mu m$ dust emission to variations in $L$.

\section{What controls the existence of a BLR?}
Some AGN do not show any broad emission lines. While there is strong evidence
that the BLR is obscured in many AGN, there are other cases
where the BLR appears to be truly missing (termed true type 2 AGN).
{\em Why do some AGN do not have a BLR?}

The distribution of observed broad line widths appears to have a cutoff at a
FWHM of about 25,000~km~s$^{-1}$ (Strateva et al. 2003), though it
is not entirely clear if this cutoff is due to a physical mechanism, or
if it is due to some observational limitation. This cutoff implies that
the BLR will not be observable in AGN with $L<L_{\rm crit}$, as explained
below. If dust sublimation sets
$R_{\rm BLR}\propto L^{0.5}$, and if gravity sets the BLR line widths
$\Delta v\propto \sqrt{M_{\rm BH}/R_{\rm BLR}}$, then these relations
imply that $\Delta v>25,000$~km~s$^{-1}$ once
$L<L_{\rm crit}=10^{25.8}(M_{\rm BH}/M_{\odot})^2$~erg~s$^{-1}$
(Laor 2003).
Thus, if the BLR is bounded from the outside by the dust sublimation radius,
and from the inside by the radius where $\Delta v>25,000$~km~s$^{-1}$,
then at $L=L_{\rm crit}$ the inner and outer boundaries merge, and no
region where the BLR can survive exists once  $L<L_{\rm crit}$. Thus all AGN
at $L<L_{\rm crit}$ should not have a BLR.

If the BLR is produced by an outflow from a dense, thin, accretion disk, then
the flow may not exist below some critical value of $L/L_{\rm Edd}$,
either because of the outflow driving mechanism properties (Nicastro 2000),
or because the disk disappears altogether and becomes a hot and dilute
thick inflow, which may occur at $L/L_{\rm Edd}\ltsim 10^{-3}-10^{-4}$.

Thus, it will be worthwhile to obtain high quality observations of
low luminosity AGN (where the host galaxy light has been carefully subtracted)
in order to 1. determine if the cutoff at $\Delta v>25,000$~km~s$^{-1}$ is
real, rather than a selection effect. 2. Determine the critical luminosity
which separates true type 2 AGN from true type 1 AGN. These observations
will provide important clues for the physical mechanisms which control
the existence of the BLR (see Laor 2003 for further discussion, including
the narrow line profiles).

\section{What controls the line strengths/ratios?}
The seminal study of Boroson \& Green (1992, hereafter BG92) established the
presence
of a set of correlations among the H$\beta$, optical Fe~II, He~II, and
[O~III]$\lambda\lambda 4959, 5007$ emission line properties. In particular
they found that a narrow
H$\beta$ line is associated with
a weak [O~III], strong optical Fe~II, and a blue
excess asymmetric H$\beta$ profile. BG92 speculated that the physical parameter
underlying these correlations (part of their eigenvector 1, or EV1) is
$L/L_{\rm Edd}$, in the sense that
objects with a narrow H$\beta$ have a high $L/L_{\rm Edd}$.

The main components of EV1 were later found to be strongly
correlated with the X-ray emission properties, such that objects with
a narrow H$\beta$ tend to have a steep soft X-ray slope.
A steep soft X-ray slope was independently suggested to be associated
with a high $L/L_{\rm Edd}$, which agreed with the
BG92 speculation. Further significant support for $L/L_{\rm Edd}$ as
the driving parameter came with the recent findings that the
H$\beta$ FWHM allows a good quantitative estimate of $M_{\rm BH}$, and thus
that a narrow H$\beta$ is generally associated with a high $L/L_{\rm Edd}$
(see Laor 2000 for a recent review).

Wills et al. (1999) extended the  BG92 correlation analysis to
the UV, and found strong correlations of UV emission line ratios with the
the EV1 optical emission line parameters. In particular, a narrow H$\beta$
is associated with weak C~IV, weak C~III], and relatively strong Si~IV+O~IV]
and N~V emission. These trends may reflect an increase in the mean BLR density
(increasing Si~III]/C~III]; Gaskell 1985; Laor et al. 1997), and metalicity
(increasing N~V/C~IV; Hamann \& Ferland 1999)
with increasing $L/L_{\rm Edd}$. Another very interesting feature is the
apparent anti-correlation between the UV Fe~II and optical Fe~II emission,
indicated by the spectral principal component analysis of Shang et al.
(2003). This effect is expected with increasing optical
depth in the Fe~II emitting gas, as an increased number of resonance scattering
of UV Fe~II photons leads to down conversion to
optical Fe~II photons. The increase in optical depth may
be due to an increase in the gas Fe abundance and overall metalicity.

A possible scenario for the above trends in the BLR properties with
$L/L_{\rm Edd}$ could be through some large scale process (interaction?)
which leads to a high rate of gas infall to the center, resulting in a
nuclear starburst, and thus metal enrichment of the gas which
continues its infall to the center, producing a high density high metalicity
BLR surrounding a high accretion rate AGN. Alternatively, Warner et al.
(2003) suggest that the BLR metalicity is driven by $M_{\rm BH}$, rather
than $L/L_{\rm Edd}$. This is supported by the known relations
between metalicity, bulge mass, and $M_{\rm BH}$, in
nearby inactive galaxies.

Further progress on the nature of the primary driver of the above
relations can be made by making direct correlations of the various
line emission properties with our best estimates of $M_{\rm BH}$ and
$L/L_{\rm Edd}$ using well defined and complete samples with high
quality optical and UV data\footnote{Note that the high ionization
lines, in particular C~IV, may not provide $M_{\rm BH}$ estimates as good as
those obtained by the Balmer lines (Baskin \& Laor, these proceedings).}.
Finally, the starburst induced high $L/L_{\rm Edd}$ scenario can be tested
through high resolution multicolor imaging with the HST of well defined
samples of nearby AGN with high and low $L/L_{\rm Edd}$.

\section{What controls the line shapes?}

The peaks of the high ionization broad emission lines are generally blueshifted
with respect to the low ionization lines, and their profiles tend to show
blue excess asymmetry, while the low ionization
lines show less asymmetry (Gaskell 1982;
Wilkes 1984, Richards et al. 2002)\footnote {Qualitatively similar
effects are seen in the narrow emission line profiles (Heckman et al. 1981,
\apj, 247, 403; Vanden Berk et al. 2001,\aj, 122, 549).}.
In a given object the line velocity shift
and asymmetry increase with the ionization level (Laor et al. 1997),
and both effects appear to be strongest in narrow H$\beta$ AGN
(Baldwin et al. 1996). Given the EV1 correlations mentioned
above, it appears that the two effects are largest in objects with a
high $L/L_{\rm Edd}$. Such a relation may be formed by the following scenario.
Radiation pressure ablates the face of the BLR ``clouds", creating a radial
outflow. As the outflow accelerates, the density drops and the ionization level
increases, creating the increased line shift and asymmetry
(the redshifted outflow is presumably
obscured). Furthermore, as $L/L_{\rm Edd}$
increases the column density of the radiatively accelerated layer increases,
creating the apparent trends with $L/L_{\rm Edd}$. Alternatively, Richards et al.
suggest the observed trends may be driven by an inclination effect.

The origin of the line asymmetry can be verified by reverberation mapping
of objects with highly asymmetric lines. If it is due to an outflow then
the blue wing should respond faster than the rest of the line. The nature of the
primary physical driver can be explored by correlating the observed
properties against $L/L_{\rm Edd}$, and by comparing the observed trends with
physical outflow models.

\pagebreak

\end{document}